\documentclass[english]{article}
\usepackage[T1]{fontenc}
\usepackage[latin9]{inputenc}
\usepackage{float}
\usepackage{amsmath}
\usepackage{graphicx}
\usepackage{setspace}
\makeatletter

\providecommand{\tabularnewline}{\\}

\usepackage{lineno} 

\usepackage[margin=1in]{geometry}

\usepackage{authblk}
\title{Supervised learning improves disease outbreak detection}
\author[1,*]{Benedikt Zacher}
\author[1]{Irina Czogiel}
\affil[1]{\small Robert Koch Institute, Department of Infectious Diseases, Berlin, Germany}
\affil[*]{\small Correspondence to: zacherb@rki.de}

\date{ }

\makeatother

\usepackage{babel}
\begin{document}
\maketitle

\vspace{0.25cm}

\begin{abstract} 
The early detection of infectious disease outbreaks is a crucial task to protect population health. To this end, public health surveillance systems have been established to systematically collect and analyse infectious disease data. A variety of statistical tools are available, which detect potential outbreaks as abberations from an expected endemic level using these data. Here, we develop the first supervised learning approach based on hidden Markov models for disease outbreak detection, which leverages data that is routinely collected within a public health surveillance system. We evaluate our model using real Salmonella and Campylobacter data, as well as simulations. In comparison to a state-of-the-art approach, which is applied in multiple European countries including Germany, our proposed model reduces the false positive rate by up to 50\% while retaining the same sensitivity. We see our supervised learning approach as a significant step to further develop machine learning applications for disease outbreak detection, which will be instrumental to improve public health surveillance systems. 
\end{abstract} 
\vspace{1cm}

\section{Introduction}

Infectious diseases are a significant threat to human health. In order
to guard against these infections, public health surveillance systems
have been established to systematically collect, analyse and interpret
data to guide public health actions \cite{Choi2012}. A central part
of public health surveillance is the early detection of disease outbreaks.
In Germany, data about notifiable infectious diseases is continuously
reported from local and federal health authorities to the Robert Koch
Institute (RKI) and collected using an electronic surveillance system
(SurvNet) for infectious disease outbreaks, which was established
in 2001 \cite{Faensen2006}. Thousands of time series of case counts
from the SurvNet database are analysed each week to detect aberrations
(i.e. potential outbreaks) from an expected endemic baseline. A plethora
of methods are available to accomplish this task, most of which use
either regression techniques or statistical process control (e.g.
\cite{enki2016,Farrington1996,hoehle2008,Noufaily2013,manitz2013}).
An excellent review of methods can be found in \cite{Unkel2012}.
\\
The Farrington algorithm is a popular method for disease outbreak
detection, which has been used in multiple European countries \cite{Hulth2010,Farrington1996}.
The RKI uses an improved version, which is implemented as 'farringtonFlexible'
(FF) in the R package surveillance \cite{Salmon2016,Hohle2007,Noufaily2013}.
In brief, the algorithm fits a quasi-Poisson Generalized Linear Model
(GLM) of the endemic baseline (i.e. no outbreaks) on past data, accounting
for possible time trends and seasonal patterns. During model fitting
the most recent weeks (default: 26) are excluded to avoid an influence
of recent outbreaks. Then the model predicts the endemic baseline
for the current week and calculates an alarm threshold either based
on the prediction interval or a quantile of the Negative Binomial
distribution. If the number of cases in the current week exceeds the
alarm threshold, a report is created for further investigation. One
crucial factor of this process is a proper balancing between the sensitivity
and the false positive rate of the generated alarms. It is important
that the recipients of the alarm reports are not overwhelmed by too
many false alarms, but at the same time, sensitivity has to be high
enough to capture relevant signals. The FF algorithm has proven to
be a good choice to accomplish this task at the RKI. In a systematic
evaluation of 19 available methods using simulations it had the lowest
false positive rate, while retaining a reasonable sensitivitiy \cite{Bedubourg2017}.
However, this algorithm still generates many false alarms in practice
that have to be verified by epidemiologists, which costs time and
money.\\
Outbreaks that are detected and confirmed e.g. by local health authorities
or the RKI, are also recorded in the SurvNet database. In this work,
we consider outbreaks that are defined according to the Protection
against Infection Act which determines compulsory registration of
'suspicion and disease of microbial foodborne disease or acute gastroenteritis
if two or more diseases of similar type occur, in which an epidemic
connection is probable or suspected' \cite{ifsg2017}. The reported
outbreaks contain information about cases, including the dates of
infection and location, which is available on the level of counties.
From this data we generate time series for counties of weekly case
counts and class labels indicating whether an outbreak occured or
not. Our proposed supervised learning approach is trained on this
data to classify weeks into an endemic (baseline) or outbreak state
by exploiting the fact that weeks with reported outbreaks exhibit
an excess number of cases compared to weeks where no outbreak was
reported. The approach is based on hidden Markov models (HMMs), which
have been used successfully in many machine learning applications
for analysis of sequential data, such as speech recognition or genome
analysis (e.g. \cite{zacher2017,Rabiner1989}). \\
The main goal of this study is to improve the prospective detection
of potential infectious disease outbreaks. In particular, we aim at
a reduction of false alarms - while maintaining sensitivity of state-of-the-art
methods - which will save valuable time during verification of the
generated alarm reports. We compare our proposed method to the FF
algorithm on simulations and real data of Salmonella and Campylobacter
infections and show that our approach reduces the false positive rate
by up to 50\% at the same sensitivity.

\section{Methods }

We assume that our data are governed by a HMM. We specify our model
using a set of $N$ time series to make use of the outbreak labels
from multiple time series during model training, i.e. at any time
point $t\in\left[1;T\right]$ in time series $n\in\left[1;N\right]$,
the corresponding obervation $o_{n,t}$ is emitted from a hidden state
$s_{n,t}\in\mathcal{K}$ that evolves over time according to a first-order
Markov process. Thus a HMM with a set of states $\mathcal{K}$ has
the following components:
\begin{itemize}
\item $\mathcal{O}=\left\{ O_{1},\ldots,O_{N}\right\} $ the sequences of
observations, where $O_{n}=(o_{n,1},\ldots,o_{n,T})$ 
\item $\mathcal{S}=\left\{ S_{1},\ldots,S_{N}\right\} $ the sequences of
latent underlying states, where $S_{n}=(s_{n,1},\ldots,s_{n,T})$
and $s_{n,t}\in\mathcal{K}$
\item $Z=\left(z_{1},\ldots,z_{T}\right)$ a sequence of covariates for
each time point
\item $\pi_{i}=\Pr(s_{n,1}=i\bigr)$ the vector of initial state probabilities,
where $\sum_{j\in\mathcal{K}}\Pr(s_{n,1}=j\bigr)=1$ 
\item $a_{ij}=\Pr\left(s_{n,t}=j\left|s_{n,t-1}=i\right.\right)$ transition
probabilities between the states $i,j\in\mathcal{K}$, \\
i.e. $\sum_{j\in\mathcal{K}}\Pr\left(s_{n,t+1}=j|s_{n,t}=i\right)=1$ 
\item $\psi_{s_{n,t}}\left(o_{n,t}\left|z_{t}\right.\right)$ is a vector
of emission functions $\psi_{s_{n,t}}\left(o_{n,t}\left|z_{t}\right.\right)=\Pr\left(o_{n,t}\left|s_{n,t},z_{t}\right.\right)$
that provides the components of the mixture distribution, i.e. the
conditional densitites of $o_{n,t}$ associated with state $s_{n,t}$,
at time $t$ in time series $n$ and covariate $z_{t}$ 
\item Parameter vector $\theta=\left(\pi_{i},a_{ij},\psi\right)$ specifies
the HMM 
\end{itemize}
In surveillance data, the observations are the reported number of
cases of a certain disease during a certain time, e.g. weeks. In our
setting, we are interested in $\mathcal{\left|K\right|}=2$ hidden
states, i.e. $s_{n,t}\in\{0,1\}$, where $s_{n,t}=1$ indicates an
ongoing outbreak with an excess number of cases at week $t$, and
$s_{n,t}=0$ applies to weeks where the case number is consistent
with the expected baseline (endemic). As many infectious diseases
and hence the corresponding surveillance data follow an annual pattern
or time trend, a natural choice for $Z$ is the sequence $z_{t}=\left(t,\cos\left(\frac{2\pi}{52}t\right),\sin\left(\frac{2\pi}{52}t\right)\right)$
to model secular and seasonal trends. We assume that the data in our
surveillance time series follows a negative Binomial distribution:
$o_{n,t}\sim\text{\ensuremath{\mathcal{NB}}}(\mu_{n,t},r_{n})$, where
$r_{n}$ is the size parameter of the negative Binomial distribution
and 
\[
\log\mu_{n,t}=\beta_{n,0}+\beta_{n,1}t+\beta_{n,2}\cos\left(\frac{2\pi}{52}t\right)+\beta_{n,3}\sin\left(\frac{2\pi}{52}t\right)+\beta_{4}s_{n,t}
\]
the $\log$ of the the expected number of cases $\mu_{n,t}$ in time
series $n$ at time $t$. Here, $\beta_{n,0}$ is the baseline number
of cases, $\beta_{n,1}$ a secular time trend and $\beta_{n,3}$ and
$\beta_{n,4}$ describe seasonal patterns for each time series $n\in\left[1;N\right]$.
The state-dependence of $o_{n,t}$ on $s_{n,t}$ - and thus the effect
of an outbreak - is incoporated by a multiplicative factor $\exp\left(\beta_{4}\right)$
that describes the excess number of cases in outbreak situations.
Note that, while $\beta_{n,0},\beta_{n,1},\beta_{n,2},\beta_{n,3}$
are specific for each time series, $\beta_{4}$ is the same for all.
This allows that information about past outbreaks is shared across
time series. This is necessary to make the model robust, since the
number of outbreaks can vary greatly between time series. If only
a few outbreaks occured in the training data of a single time series,
fitting a model with a specific parameter for the effect of an outbreak
would not generalize well on new data. In particular, it would not
be possible to fit a model on a single time series if there is no
outbreak in the past training data. 

\paragraph*{Parameter learning.}

Since there is a reporting delay of outbreaks - i.e. outbreaks in
the recent past are not yet recorded in SurvNet - we exclude $u$
time units from our training data: $T_{train}=T-u$. Assuming independence
between individual time series, the likelihood of our model with known
state sequences $S_{n}^{train}=\left(s_{n,1},\ldots,s_{n,T_{train}}\right)\in\mathcal{S}_{train}$
and observation sequences $O_{n}^{train}=\left(o_{n,1},\ldots,o_{n,T_{train}}\right)\in\mathcal{O}_{train}$
is: 
\begin{align*}
\mbox{Pr}(\mathcal{O}_{train},\mathcal{S}_{train}\left|\theta,Z\right.) & =\prod_{n=1}^{N}\mbox{Pr}(O_{n}^{train}\mid S_{n}^{train},Z,\theta)\cdot\mbox{Pr}(S_{n}^{train}\left|\theta\right.)\\
 & =\prod_{n=1}^{N}\left[\prod_{t=1}^{T_{train}}\psi_{s_{n,t}}(o_{n,t}\left|z_{t}\right.)\cdot\prod_{t=2}^{T_{train}}a_{s_{n,t-1}s_{n,t}}\cdot\pi_{s_{n,1}}\right]
\end{align*}

Maximum likelihood estimation of model parameters $\pi_{i}$ and $a_{i,j}$
is straightforward:

\begin{align*}
a_{i,j} & =\frac{\sum_{n=1}^{N}\sum_{t=2}^{T_{train}}\delta\left(s_{n,t-1},i\right)\delta\left(s_{n,t},j\right)}{\sum_{n=1}^{N}\sum_{t=2}^{T_{train}}\delta\left(s_{n,t-1},i\right)}\\
\pi_{i} & =\frac{\sum_{n=1}^{N}\delta\left(s_{1},i\right)}{N}
\end{align*}

where $\delta(i,j)=\left\{ \begin{tabular}{l}
 1, if i = j \\
 0, if i \ensuremath{\neq} j 
\end{tabular}\right.$. Estimation of $\beta=\left(\beta_{n,0},\beta_{n,1},\beta_{n,2},\beta_{n,3},\beta_{4}\right)$
is carried out using the Iteratively Reweighted Least Squares algorithm
for Generalized Linear Models, for which the R functions glm.fit()
and glm.nb() are used \cite{Venables2002,Ihaka1996}. 

\paragraph*{Posterior probability of an outbreak.}

In order to determine whether time point $T$ in time series $n$
is in the endemic or in the outbreak state, the posterior probability
\[
\Pr\left(s_{n,T}=1\left|O_{n},\theta\right.\right)=\frac{\Pr\left(s_{n,T}=1,O_{n}\left|\theta\right.\right)}{\sum_{s_{n,T\in\left\{ 0,1\right\} }}\Pr\left(s_{n,T},O_{n}\left|\theta\right.\right)}
\]
 is calculated. This can be done efficiently using recursive computation
of the forward probabilites $\alpha_{n,t}\left(i\right)=\Pr\left(s_{n,t}=i,O_{n}\left|\theta\right.\right)$
of a HMM \cite{Rabiner1989}.

\paragraph*{Data and model fitting.}

Time series data were extracted for Salmonella and Campylobacter infections
from the SurvNet database (accessible online: https://survstat.rki.de/),
which collects reports about notifiable diseases at the RKI. Data
were aggregated by disease, local health authorities - representing
counties or districts in Germany - and weeks. Weekly outbreak labels
were assigned to counties if at least two cases were part of an outbreak
in that week. For a further description of the outbreak data and labels
see \cite{Ghozzi2019}. Time series were randomly assigned to 20 equally
sized groups, ensuring that each group had enough outbreaks for training.
For each week in 2010-2017 models were trained on data using the past
five years excluding the latest 26 weeks.

\paragraph*{Simulations.}

To assess model performance in a controlled setting, we adapted 14
different simulation scenarios as proposed in Noufaily et al. \cite{Noufaily2013}.
In short, expected case counts for each time series with a length
of 624 were simulated from a linear model including Fourier terms
for an annual seasonal pattern and an optional time trend: $\mu_{t}=\exp\left(\beta_{0}+\beta_{1}t+\beta_{2}\cos\left(\frac{2\pi}{52}t\right)+\beta_{3}\sin\left(\frac{2\pi}{52}t\right)\right)$.
Parameters for all simulation scenarios are depicted in Supplementary
Table 1. The state sequences of endemic and outbreak weeks were simulated
using a transition matrix, where for each time series, $a_{00}$ was
sampled from a uniform distribution from the interval $\left[0.9;1\right]$
and $a_{11}$ was sampled from $\left[0.4;0.6\right]$. As an additional
parameter, each scenario is assigned a dispersion parameter $\phi\geq1$
(Supplementary Table 1). Weekly case counts of the endemic state were
then sampled from a Negative Binomial with mean $\mu_{t}$ and variance
$\phi\mu_{t}$. For time points in the outbreak state, $\mu_{t}$
was chosen such that the power of detecting the outbreak with a p-value
< 0.01 from the endemic distribution was 0.5.

\paragraph*{Evaluation and benchmarking.}

Performance of the HMM was compared to the FF algorithm, which is
currently the method of choice for outbreak detection at the RKI.
The farringtonFlexible() function from the R package surveillance
was used with the following (default) control parameters: noPeriods
= 10, b = 5, w = 3, weightsThreshold = 2.58, pastWeeksNotIncluded
= 26, thresholdMethod = \textquotedbl nbPlugin\textquotedbl , alpha
= 0.01 \cite{Hohle2007}. ROC curves, sensitivity, false positive
rate and precision were computed using the R package ROCR \cite{Sing2005}.

\paragraph*{Availability. }

The R source code of the model is part of this paper as Supplementary
Information.

\section{Results}

We briefly want to illustrate the workflow and components of our method
with an application to a set of infectious disease surveilance data
using a simulated example of the twelve districts of Berlin (Figure
1). By default the HMM uses five years of training data for outbreak
detection in the current week, however other durations are possible.
The past 26 weeks are excluded from the training data to avoid model
training on incomplete data due to possible reporting delays. Since
the frequency of disease outbreaks varies between counties, one model
is trained on multiple time series (in our example one model is trained
for the 12 districts of Berlin) to make sure that enough past outbreaks
are available for training (Figure 1A). The fitted HMM consists of
a linear predictor, initial state and transition probabilities (Figure
2B). Using the generalized linear model, the expected number of cases
of the endemic and the outbreak state is predicted for the current
week. Then the probability of an outbreak in the current week is calculated
for all time series, which is depicted on a map in our example (Figure
3C). An alarm will be triggered, if the probability exceeds a chosen
threshold.\\
We applied the HMM and the FF algorithm to Salmonella and Campylobacter
cases reported from more than 400 counties in Germany and simulated
data (Methods). Figure 2 shows Salmonella and Campylobacter data aggregated
by week for Germany. The number of infections and outbreaks per week
in Germany show a strong seasonal pattern and there is a decrease
of Salmonella cases and a low increase of Campylobacter cases over
time. This justifies the choice of our model to include seasonal and
secular trends. We applied the models to predict outbreaks from 2010
to 2017. During this time there were 2126 Salmonanella outbreaks (according
to our definiton, see Methods) with a duration of 1-8 weeks and 2260
Campylobacter outbreaks with a duration of 1-16 weeks. The average
number of cases in counties which reported an outbreak in a week exhibits
a marked increase compared to the average cases reported by counties
where no outbreak occured in that week. This shows that weekly case
counts of reported outbreaks are well separated from endemic weeks
and therefore might be a valuable source of information to improve
outbreak detection. \\
In all our applications the HMM consistently outperformed the FarringtonFlexible
algorithm (Figure 3). At the same sensititvity the HMM had a lower
false positive rate and a higher precision than the FF algorithm.
To further investigate the different performances we set alarms with
the FF algorithm using the 'nbPlugin' threshold with $\alpha=0.01$,
which are the default settings currently used at the RKI (Figure 3
A,B,C). Further, we chose the cutoff for the posterior probability
of an outbreak such that the sensitivity was the same as for FF. This
resulted in a sensitivity of 0.27 for simulated, 0.21 for Salmonella
and 0.07 for Campylobacter data. At the same sensitivtiy, the false
positive rate (fpr) for the HMM (0.0016) was roughly 50\% lower than
for FarringtonFlexible (0.0034), and the precision increased from
0.86 (FF) to 0.93 (HMM) on simulated data. On Salmonella data the
fpr was 0.0074 for HMM, 0.0085 for FF and precisions amounted to 0.50
and 0.46 respectively. On the Campylobacter data, fpr was reduced
roughly by 40\% (HMM: 0.0053, FF: 0.009) and precision increased from
0.14 (FF) to 0.21 (HMM). \\
Furthermore, we investigated the distribution of posterior probabilities
of the outbreak state in endemic weeks and weeks with reported outbreaks
(Figure 3 G,H,I). There is a strong increase in posterior probabilities
in outbreak weeks compared to endemic weeks in all scenarios. Moreover
the posterior probability of an outbreak also increased with the size
of reported outbreaks. Posterior probabilites in reported outbreak
weeks were generally higher for Salmonella than Campylobacter data.
This matches the fitted outbreak effects ($\exp\left(\beta_{4}\right)$,
see Methods) of our HMMs. The average increase in the number of cases
during an outbreak ranged from 1,7 - 8,3 fold (mean: 3.6) for Salmonella
and 1,3 - 2,8 fold (mean: 1,8) for Campylobacter. \\
We also calculated absolute numbers of alarms and their overlap between
methods and with reported outbreaks from the SurvNet database (Figure
3 J,K,L). Despite a significant overlap of correctly recalled outbreak
weeks for both methods across the three applications, there is a varying
amount of outbreak weeks that are exclusively recalled by one of the
methods. For instance for Campylobacter, 190 reported outbreaks are
recalled by each method. However, among those each method identifies
81 outbreaks that are not detected by the respective other approach. 

\section{Discussion}

We introduced a supervised learning approach using hidden Markov models,
which significantly improved performance of outbreak detection on
simulated data, and real Salmonella and Campylobacter data. In our
comparison with the FF algorithm, we showed that the HMM produced
significantly less false alarms. Thus the application of our method
in practice could reduce the workload of epidemiologists and save
time and money. To the best of our knowledge, this is the first study
that leverages data of past reported outbreaks from a surveillance
system for their prospective detection. \\
The use of HMMs for modeling of epidemiological time series was also
proposed in previous studies \cite{lestrat1999,Watkins2009,Pelat2017,Rath2003,Beneito2008}.
However, these were focused on the unsupervised segmentation of infectious
disease data. The unsupervised HMMs showed good performance e.g. in
the segmentation of Influenza cases lacking covariates for seasonal
and periodic patterns. However, when using unsupervised HMMs with
periodic and seasonal terms as proposed in \cite{lestrat1999} for
the Campylbacter and Salmonella data, they did not perform as well
as the FF algorithm. Thus we did not include them in the performance
evaluation of this work.\\
It is also important do discuss some limitations of our approach.
One obvious caveat of any supervised learning approach is, that outbreak
labels are needed for model training. This data might not be available
in other surveillance systems or might not be collected for all diseases
of interest. In such cases one could either try to label time series
afterwards or resort to other well established algorithms such as
the FF algorithm. \\
The performance of a supervised learning approach for outbreak detection
also depends on the quality of the outbreak labels. Our model exploits
the fact that outbreak weeks have a higher average number of cases
than endemic weeks. However, weeks with an excess number of cases
are not reported (and labeled) as outbreaks if the compulsory registration
criteria defined in the Protection against Infection Act are not met.
This is not problematic for our approach as long as reported outbreak
weeks show an increased average number of cases, separating outbreak
from endemic weeks. This is the case for average case counts aggregated
for weekly endemic and epidemic weeks for Germany (Figure 2). It is
also verified by the fitted models, since the 'outbreak effect' parameters
$\exp\left(\beta_{4}\right)$ show a strong average increase in the
number of cases in outbreak weeks. Another issue might be that smaller
outbreaks are easily overlooked and thus not labeled. Apparently,
small outbreaks are not very well distinguished from the endemic level,
which is supported by the correlation between outbreak size and the
assigned probability of an outbreak by the HMM (Figure 3 G,H,I). 73\%
(n=1915) of Salmonella and 83\% (n=2271) of Campylobacter outbreak
weeks exhibit outbreaks with only 2 or 3 cases. This also explains
the low sensitivity of both methods, especially for the Campylobacter
data set. Thus unlabeled small outbreaks are not problematic for our
approach since they are not well distinguished from the endemic level.
Ultimately the use of outbreak labels as defined in this study is
justified by the significant improvement in the practical application
to Salmonella and Campylobacter data. \\
Another limitation is that the assumption of (conditional) independence
of observed time points in modeling infectious disease surveillance
data is questionable, since the number of infections from one week
might affect the next week. Although our proposed HMM does not take
into account the dependence of subsequent observations, it incorporates
dependence of the state of a current week (outbreak or endemic) on
the previous week. \\
Future efforts will be needed to prove application of our proposed
approach in daily practice of infectious disease surveillance. However,
our results are promising that leveraging outbreak data with supervised
learning will improve disease outbreak detection. Thus we foresee
our approach to be instrumental to improve public health surveillance
systems in the future.

\section*{Acknowledgments}

We thank Stéphane Ghozzi for fruitful discussions which lead to the
idea of using outbreak labels for the development of a supervised
learning algorithm and Alexander Ullrich for help with data extraction,
preprocessing and feedback on the manuscript. BZ was supported by
BMBF (Medical Informatics Initiative: HIGHmed) and the collaborative
management platform for detection and analyses of (re-) emerging and
foodborne outbreaks in Europe (COMPARE: European Union's Horizon 2020
research and innovation programme, grant agreement No. 643476).

\section*{Author Contributions}

BZ initiated the study, developed the method, carried out all analyses
and wrote the manuscript. IC contributed to method development and
writing of the manuscript. Both authors read an approved the final
version of the manuscript.

\section*{Conflict of interest}

The authors declare that they have no conflict of interest. 

\bibliographystyle{unsrt}
\bibliography{bibl}

\section*{\newpage Figures}

\begin{figure}[H]
\includegraphics[scale=0.7]{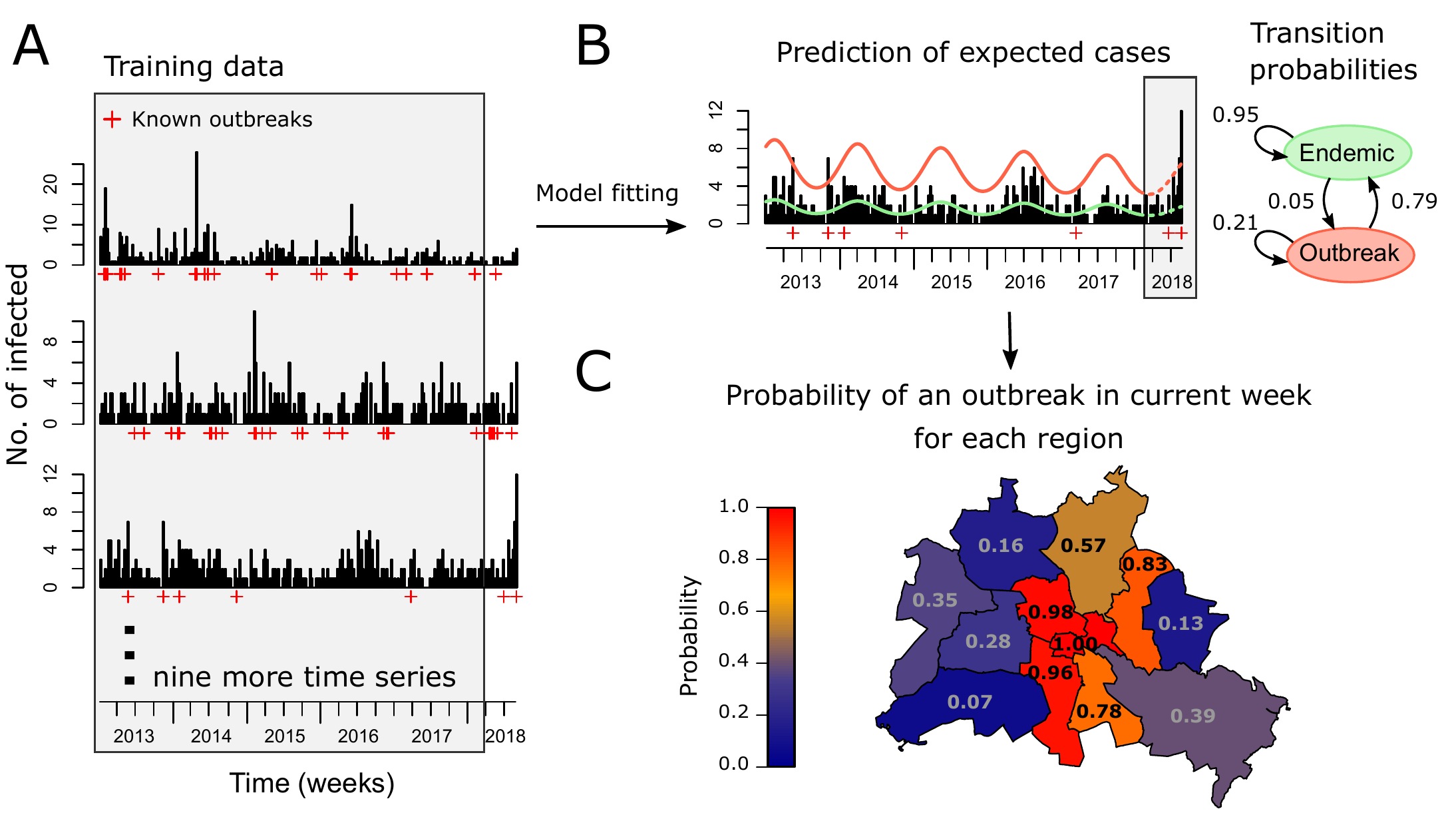}\caption{Overview of the method using simulated data for twelve districts in
Berlin. For reasons of data protection, outbreak data cannot be shown
for individual counties. To get a realistic sceanrio for illustration
of the method, the data was simulated from HMMs which were fit to
real data of Salmonella infections. (A) The example shows five years
of data of simulated Salmonella infections in Berlin. The hidden Markov
model is trained on reported cases and outbreaks, indicated by the
shaded area. (B) The expected number of cases in the endemic (green)
and the outbreak (red) state are shown, which are extrapolated up
to the current week (dashed lines). The fitted transition proababilities
are shown as a graph. (C) For each region, the posterior probabilities
are calculated.}
\end{figure}

\begin{figure}[H]
\includegraphics[scale=0.7]{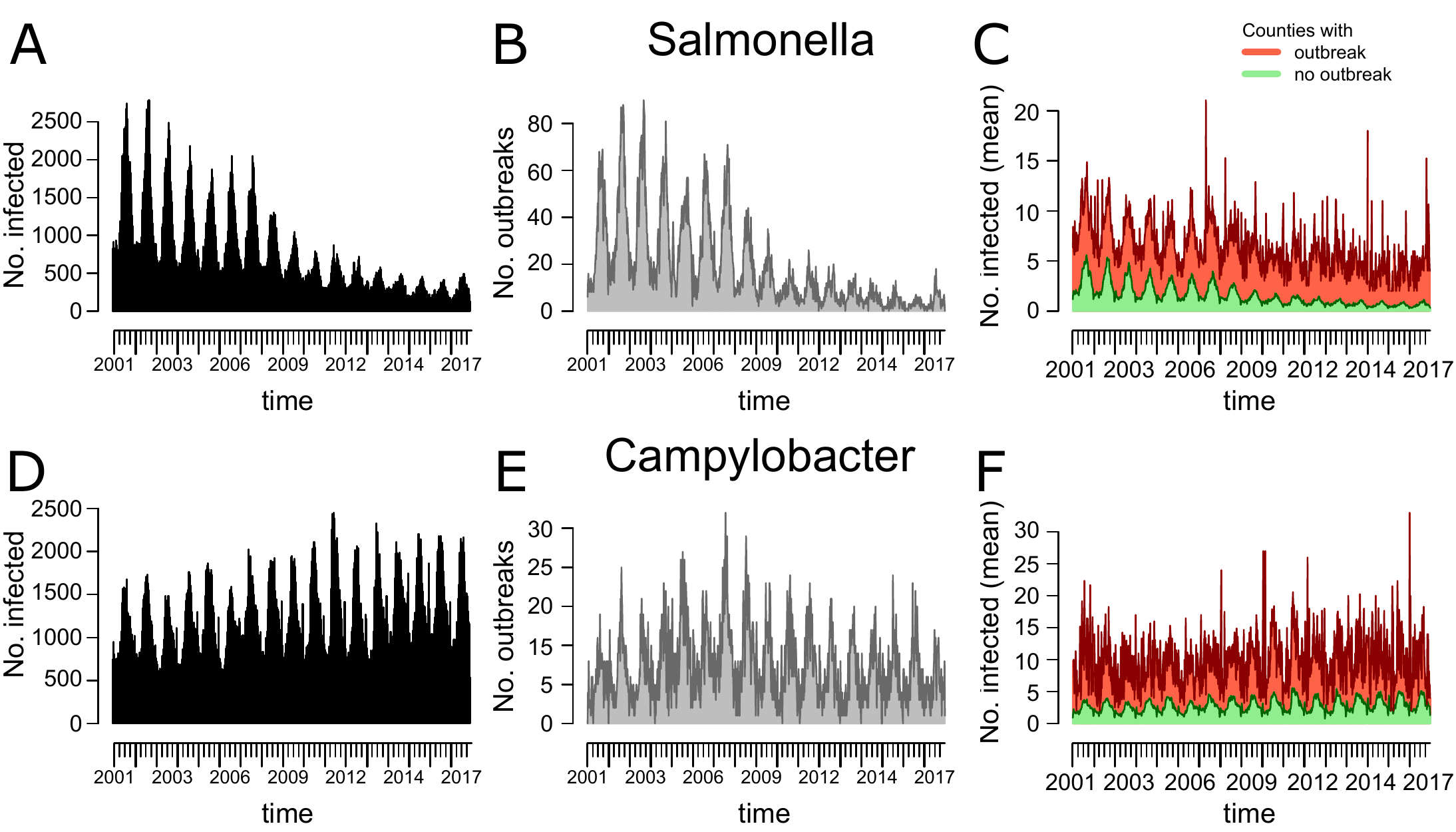}

\caption{Aggregated weekly data of Salmonella and Campylobacter infections
in Germany reported to the Robert Koch Institute. (A) The number of
Salmonella infections and (B) the number of counties with a reported
Salmonella outbreak show a seasonal and secular pattern. (C) The mean
number of cases in counties with a reported Salmonella outbreak (red)
is higher than the mean number of cases in counties without outbreak
(green). (D-F) The same as (A-C) for Campylobacter infections.}
\end{figure}

\begin{figure}[H]
\includegraphics[scale=0.65]{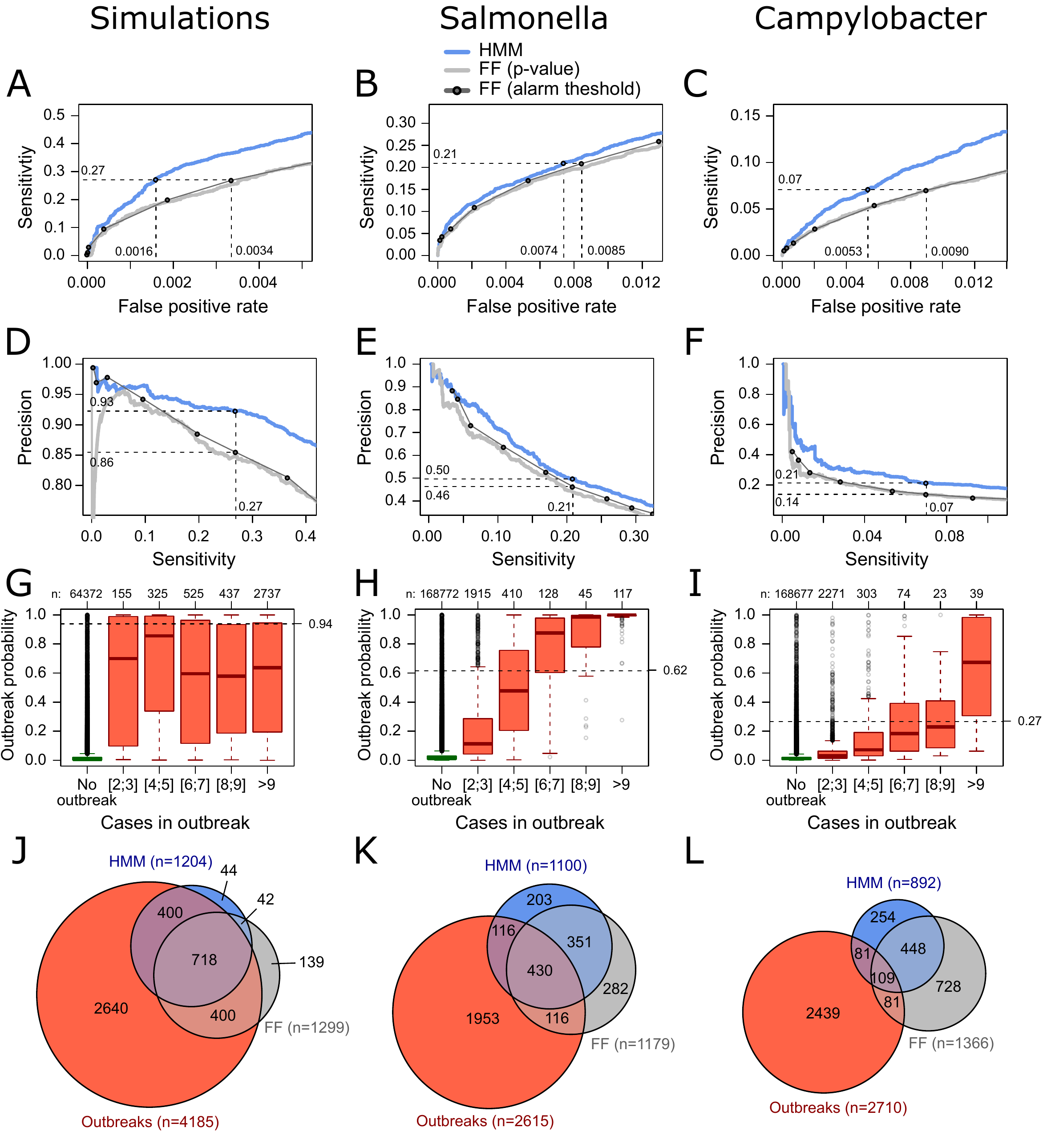}

\caption{Results of the hidden Markov model (HMM) and the FarringtonFlexible
(FF) algorithm are shown for simulations (A,D,G,J), Salmonella (B,E,H,K)
and Campylobacter (C,F,I,L). Models were applied to data aggregated
by county and week from 2010-2017 for Salmonella and Campylobacter.
(A,B,C) False postive rate is plotted against sensitivity for posterior
probability of the outbreak state (HMM) and 1 - p-value (solid grey
line) for the FF algorithm. Dashed lines and grey points show sensitivity
and false positive rate with cutoffs $10^{-6}$, $10^{-5}$, $10^{-4}$,
$0.001$, $0.005$, $0.01$ and $0.02$ for the FF algorithm with
threshold method 'nbPlugin'. (D,E,F) Sensitivity is plotted against
precision for posterior probability of the outbreak state (HMM) and
1 - p-value for the FF algorithm. Dashed lines and grey points show
sensitivity and false positive rate with chosen cutoffs as in (A-C).
(G,H,I) Boxplots of posterior probabilities in endemic weeks (green)
and weeks with reported outbreaks (red) are shown. Outbreaks were
further divided by their size (i.e. the number of cases reported in
the respective outbreaks). For simulations the cases in an outbreak
were chosen as the excess number of cases compared to the expected
endemic level. Dashed lines indicate the cutoff for posterior probabilites.
(J,K,L) Venn diagrams showing the overlap of predicted outbreak weeks
of the HMM and FF algorithm with reported outbreaks.}
\end{figure}

\section*{Supplementary Information}

\begin{table}[H]
\begin{centering}
\begin{tabular}{|c|c|c|c|c|c|}
\hline 
Scenario & $\beta_{0}$ & $\beta_{1}$ & $\beta_{2}$ & $\beta_{3}$ & $\phi$\tabularnewline
\hline 
\hline 
1 & 0.1 & 0 & 0.6 & 0.6 & 1.5\tabularnewline
\hline 
2 & 0.1 & 0.0025 & 0.6 & 0.6 & 1.5\tabularnewline
\hline 
3 & -2 & 0 & 0.1 & 0.3 & 2\tabularnewline
\hline 
4 & -2 & 0.005 & 0.1 & 0.3 & 2\tabularnewline
\hline 
5 & 1.5 & 0 & 0.2 & -0.4 & 1\tabularnewline
\hline 
6 & 1.5 & 0.003 & 0.2 & -0.4 & 1\tabularnewline
\hline 
7 & 0.5 & 0 & 0.5 & 0.5 & 5\tabularnewline
\hline 
8 & 0.5 & 0.002 & 0.5 & 0.5 & 5\tabularnewline
\hline 
9 & 2.5 & 0 & 1 & 0.1 & 3\tabularnewline
\hline 
10 & 2.5 & 0.001 & 1 & 0.1 & 3\tabularnewline
\hline 
11 & 3.75 & 0 & 0.1 & -0.1 & 1.1\tabularnewline
\hline 
12 & 3.75 & 0.001 & 0.1 & -0.1 & 1.1\tabularnewline
\hline 
13 & 5 & 0 & 0.05 & 0.01 & 1.2\tabularnewline
\hline 
14 & 5 & 0.0001 & 0.05 & 0.01 & 1.2\tabularnewline
\hline 
\end{tabular}\caption{Parameters for all 14 simulation scenarios are shown.}
\par\end{centering}
\end{table}

\end{document}